\documentclass[prl,twocolumn,showpacs,letterpaper,showpacs,superscriptaddress]{revtex4}
\usepackage{graphicx,amsmath,amssymb,amsfonts,latexsym,color,dcolumn,bm,epsfig,subfigure}

\begin{document}

\title{Casimir-Lifshitz Theory and Metamaterials}

\author{F. S. S. Rosa}
\affiliation{Theoretical Division, Los Alamos National Laboratory, Los Alamos, NM 87545, USA}

\author{D. A. R. Dalvit}
\affiliation{Theoretical Division, Los Alamos National Laboratory, Los Alamos, NM 87545, USA}

\author{P. W. Milonni}
\affiliation{Theoretical Division, Los Alamos National Laboratory, Los Alamos, NM 87545, USA} 

\date{\today}

\begin{abstract}
Based on a generalization of the Lifshiftz theory we calculate Casimir forces involving magnetodielectric and possibly anisotropic metamaterials, focusing on the possibility of repulsive forces. 
It is found that Casimir repulsion decreases with magnetic dissipation, and even a small Drude background
in metallic-based metamaterials acts to make attractive a Casimir force that would otherwise be predicted to be repulsive. The sign of the force also depends sensitively on the degree of optical
anisotropy of the metamaterial, and on the form of the frequency-dependency of the magnetic response.

\end{abstract}


\maketitle

The growing importance of both Casimir effects \cite{reviewCasimir} and metamaterials \cite{reviewMM} raises questions concerning
the nature of Casimir forces between metamaterials. The rather complicated but highly successful Lifshitz 
theory \cite{lif} of Casimir forces assumes isotropic and nonmagnetic media. Calculations based on its generalization to magnetodielectric media suggest that Casimir forces involving metamaterials could have some very interesting properties, e.g., left-handed metamaterials might lead to
repulsive Casimir forces (``quantum levitation") \cite{henkel,ulf}. Metamaterials, however, typically have narrow-band magnetic response and are anisotropic, and so questions naturally arise concerning the validity of such predictions 
for real metamaterials. Such questions are addressed here.

The Lifshitz theory, as in all of macroscopic QED, is based on the assumption that the media can be approximated as continua described
by permittivities and permeabilities. This requires that field wavelengths 
contributing significantly to the force be large compared to length scales characterizing the metamaterial structures
and separations. The original Lifshitz formula is easily generalized to the case of magnetically permeable, anisotropic media. One approach is based on the general scattering formula describing the Casimir interaction energy between two bodies in vacuum. Consider media 1 and 2 occupying the half-spaces $z<0$ and $z>d$, respectively, and separated by vacuum (region 3). The zero-temperature Casimir interaction energy is given by $E(d) = (\hbar / 2 \pi) \int_0^{\infty} d\xi \log {\rm det}\, {\cal D}$, where ${\cal D} = 1 - {\cal R}_1 e^{-{\cal K} d} {\cal R}_2 e^{-{\cal K} d}$ \cite{kats}. Here ${\cal R}_{1,2}$ are reflection operators at the interfaces $z=0$ and $z=d$, respectively, and $e^{-{\cal K} d}$ represents a one-way propagation between the two media. Assuming the media are spatially homogeneous, only specular reflection occurs, and the resulting Casimir force per unit area $A$ between the two media is
\begin{equation}
\label{eq1}
\frac{F(d)}{A}= 2 \hbar \int_0^{\infty} \frac{d \xi}{2 \pi} \int \frac{d^2 {\bf k}_{\|}}{(2 \pi)^2}
\, K_3 \, {\rm Tr} \frac{{\bf R}_1 \cdot {\bf R}_2 e^{-2K_3 d} }{
1-{\bf R}_1 \cdot {\bf R}_2 e^{-2K_3 d}} ,
\end{equation}
where ${\bf R}_{1,2}$ are the $2\times 2$ reflection matrices, 
${\bf k}_{\|}$ is the transverse wavevector, and $K_3 = \sqrt{k_{\|}^2 + \xi^2/c^2}$. 
A positive (negative) value of the force corresponds to
attraction (repulsion). Note that, based on the well-known analytic properties of the permittivities and permeabilities, the reflection matrices here are evaluated at imaginary frequencies $\omega=i \xi$.  
For general anisotropic media these reflection matrices are defined as
\begin{eqnarray}
\label{ReflectionMatrices}
{\bf R}_j = \left[
\begin{array}{cc}
   r^{ss}_j (i \xi, {\bf k}_{\|}) &  r^{sp}_j (i \xi, {\bf k}_{\|}) \\
   r^{ps}_j (i \xi, {\bf k}_{\|}) &  r^{pp}_j (i \xi, {\bf k}_{\|}) 
\end{array} \right]
\;\;\;\; (j=1,2),
\end{eqnarray} 
where $r^{ab}_j$ is the ratio of a reflected field with $b$-polarization divided by an incoming field with $a$-polarization, the indices $s,p$ corresponding respectively to perpendicular and parallel polarizations with respect to the plane of incidence. In the case of isotropic media the off-diagonal elements vanish, the diagonal elements are given by the usual Fresnel expressions in terms of the electric
permittivity $\epsilon(i \xi)$ and magnetic permeability
$\mu(i \xi)$, and Eq.(\ref{eq1}) then leads to the usual Lifshitz formula for the force \cite{lif}.
 
One possible route towards Casimir repulsion invokes non-trivial magnetic
permeability \cite{other}. Casimir repulsion was predicted by Boyer \cite{boyer} for perfectly conducting and perfectly permeable parallel plates, but it may
also occur between real plates as long as one is mainly (or purely) non-magnetic and the other mainly (or purely) magnetic \cite{klich}.
The latter possibility has been considered unphysical 
\cite{capassocomment}, since naturally occurring materials do not show strong magnetic response at near-infrared/optical frequencies (corresponding to gaps $d=0.1-10$ $\mu$m, for which Casimir forces are typically measured), which is usually assumed necessary to realize repulsion (but see below).
However, recent progress in nano-fabrication has resulted in 
metamaterials with magnetic response in the visible range of the
spectrum \cite{opticalMM}, fueling the hope for quantum levitation
\cite{henkel,ulf}. In the following we discuss these
expectations and consider key roadblocks in the pursuit of 
metamaterials-based Casimir repulsion.

%%%%%%%%%%%%%%%%%%%%%%%%%%%%

{\it Quantum levitation and metamaterials---}
It has been shown that a perfect-lens slab
sandwiched between two perfect conducting planar plates
implies a Casimir repulsion between the two
plates \cite{ulf}. The condition $\epsilon(\omega)=\mu(\omega)=-1$
on which this interesting prediction is based is  
incompatible with the  Kramers-Kronig relations for
causal, passive materials (${\rm Im} \epsilon(\omega)>0$
and ${\rm Im} \mu(\omega)>0$), which do not allow 
$\epsilon(\omega)=\mu(\omega)=-1$ at all frequencies. 
Of course we can have the real parts of $\epsilon(\omega)$ and $\mu(\omega)$
approximately equal to $-1$ near a given resonance 
frequency of the metamaterial (with non-zero imaginary parts),  
but this is not a necessary condition for obtaining
repulsion. Given that the integral in Eq.(1) is dominated
by low-frequency modes $\xi < c/d$, it 
implies that a repulsive force is in principle possible for a passive
left-handed medium as long as the low-frequency magnetic response $\mu(i \xi)$ along
frequencies $\xi$ is sufficiently larger than 
$\epsilon(i \xi)$. In such a situation
the repulsion is a consequence of the low-frequency behavior of $\epsilon(i \xi)$
and $\mu(i \xi)$ and not of the fact that the medium happens to be left-handed
in a narrow band about some real resonant frequency.

An alternative scenario was also considered in \cite{ulf}, where it was argued from the
Lifshitz formula that Casimir repulsion can occur with active (amplifying) 
metamaterials, satisfying the perfect-lens condition $\epsilon(i \xi)=\mu(i \xi)=-1$
at frequencies $\xi < c/d$. In our view the use of the Lifshitz formalism 
for active materials requires further analysis to account, for instance, for the 
inevitable amplification of noise in active media, which is not taken into account in the Lifshitz
formula (\ref{eq1})\cite{activeMM}.

%%%%%%%%%%%%%%%%%

{\it Repulsion via metallic-based metamaterials ---} 
Application of the Lifshitz formula requires the knowledge of $\epsilon_j(i \xi)$ and $\mu_j(i \xi)$ for a large range of frequencies, up to the order of $\xi \simeq c/d$.
Such functions can be evaluated via the Kramers-Kronig (KK)
relations in terms of optical responses 
$\epsilon_j(\omega)$ and $\mu_j(\omega)$ at real frequencies. The KK relations imply that dispersion data 
are required over a large range of frequencies $\omega$,
typically in the low-frequency range for metals. This point
is very important, as it shows that knowledge of the optical response
of a metallic-based metamaterial near a resonance is not sufficient for the 
computation of Casimir forces: the main contribution
to $\epsilon(i \xi)$ and $\mu(i \xi)$ typically comes
from frequencies lower than the resonance frequency. 
Although somewhat counterintuitive, this fact explains the recent experiment
[16] in which it was found that a sharp change of the optical
response  of a hydrogen-switchable mirror at optical frequencies $\omega \simeq c/d$ did not appreciably change the Casimir force. 
The fact that the dominant contribution to the force comes from
low frequencies over bandwidths wide compared to typical metamaterial resonances also implies
that repulsive forces are in principle possible without the
requirement that the metamaterial resonance should be near the frequency scale defined by the
inverse of the gap $d$ of the Casimir cavity.

\begin{figure}
\begin{center}
\hspace{-20pt}
\scalebox{0.3}{\includegraphics{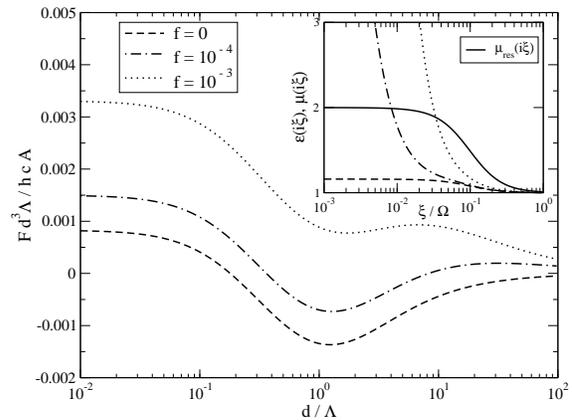}}
\vspace{-7pt}
\caption{Casimir force between a gold half-space and a silver-based planar metamaterial for 
different filling factors. The frequency scale $\Omega=2 \pi c / \Lambda$ is chosen as the silver
plasma frequency $\Omega_D=1.37 \times 10^{16}$ rad/sec. Parameters are: for the metal, $\Omega_1/\Omega=0.96$, $\gamma_1/\Omega=0.004$, and for the metamaterial, 
$\Omega_D/\Omega=1$, $\gamma_D/\Omega=0.006$, 
$\Omega_e/\Omega=0.04$, $\Omega_m/\Omega=0.1$, 
$\omega_e/\Omega=\omega_m/\Omega=0.1$, 
$\gamma_e/\Omega=\gamma_m/\Omega=0.005$.
The inset shows the magnetic permeability $\mu_{\rm res}(i \xi)$
and the electric permittivity $\epsilon_2(i \xi)$ for the different filling factors.}
\label{Comparison2}
\end{center}
\end{figure} 

Let us consider the simple example of a metallic half-space 1
in front of a metallic metamaterial 2. For the metal we assume the usual Drude model,
$\epsilon_1(\omega) = 1 - \Omega_1^2/(\omega^2 + i \gamma_1 \omega) \; ; \; \mu_1(\omega) = 1$ ,
where $\Omega_1$ is its plasma frequency and $\gamma_1$ the dissipation coefficient.
Typical metamaterials have resonant electromagnetic response at certain frequencies;
in the simplest description, the  resonant behavior can be modelled by a Drude-Lorentz model,
\begin{equation}
\label{Drude-Lorentz}
\epsilon_{\rm res}(\omega) \, , \, \mu_{\rm res}(\omega) = 1 - \frac{\Omega_{e,m}^2}{\omega^2 - 
\omega_{e,m}^2 + i \gamma_{e,m} \omega} ,
\end{equation}
which $\omega_{e}$ ($\omega_{m}$) is the electric (magnetic) resonance frequency, and
$\gamma_e$ ($\gamma_m$) is the metamaterial electric (magnetic) dissipation parameter.
For metamaterials that are partially metallic, such as split-ring resonators (operating in the GHz-THz range) and fishnet arrays (operating
in the near-infrared/optical) away from resonance, it is reasonable to assume that the dielectric function has a Drude background response in addition to the resonant part. For frequencies $\omega=i \xi$, the total
dielectric permittivity of half-space 2 can then be modeled as
\begin{equation}
\label{Drude Background}
\epsilon_2(i \xi)  = 1 + f \frac{\Omega_{D}^2}{\xi^2 +  \gamma_{D}\xi}  + (1-f) \frac{\Omega_{e}^2}{\xi^2 + \omega_{e}^2 + \gamma_{e}\xi} ,
\end{equation}
where $f$ is a filling factor that accounts for the fraction of metallic structure contained in the metamaterial, and $\Omega_D$, $\gamma_D$ are the Drude parameters of this metallic structure. The
total magnetic permeability is given by the resonant part
alone, Eq. (\ref{Drude-Lorentz}).
As the Drude background clearly overwhelms the resonant contribution $\epsilon_{\rm res}$ for low 
frequencies $\xi$, it  contributes substantially to the Lifshitz force. In Fig. \ref{Comparison2} we plot the Casimir-Lifshitz force between a metallic half-space and a metallic-based planar metamaterial for different filling factors
at zero temperature. (The effect of finite temperature is to
decrease the amount of repulsion \cite{henkel}). Without the Drude contribution ($f=0$) there is repulsion for a certain range of distances, as long as half-space 2 is mainly magnetic 
($\epsilon_2(i \xi) < \mu_2(i \xi)$). However, even a small amount of metallic background ($f>0$) can spoil the possibility of repulsion for any separation (see also \cite{irina}). In view of these observations, metamaterials with  negligible Drude background, including those based on
polaritonic photonic crystals \cite{photoniccrystals} or
on dielectric structures \cite{dielectrics} showing magnetic
response, seem more likely candidates for Casimir repulsion.

Given that dissipation plays an important role in metallic-based metamaterials, especially
those operating at high frequencies, we conclude this section by noting that, as a general rule, dissipation tends to reduce Casimir repulsion. 
For example, increasing the ratios $\gamma_e/\Omega_e=\gamma_m/\Omega_m$ from $0.01$ to $1$
decreases the magnitude of Casimir repulsion by approximately $50\%$ at $d/\Lambda=1$ (for
$f=10^{-4}$, $\Omega_e/\Omega=0.04$, $\Omega_m/\Omega=0.1$, and all other parameters as in Fig. 1).

\begin{figure}[t]
\hspace{-10pt}
\scalebox{0.28}{\includegraphics{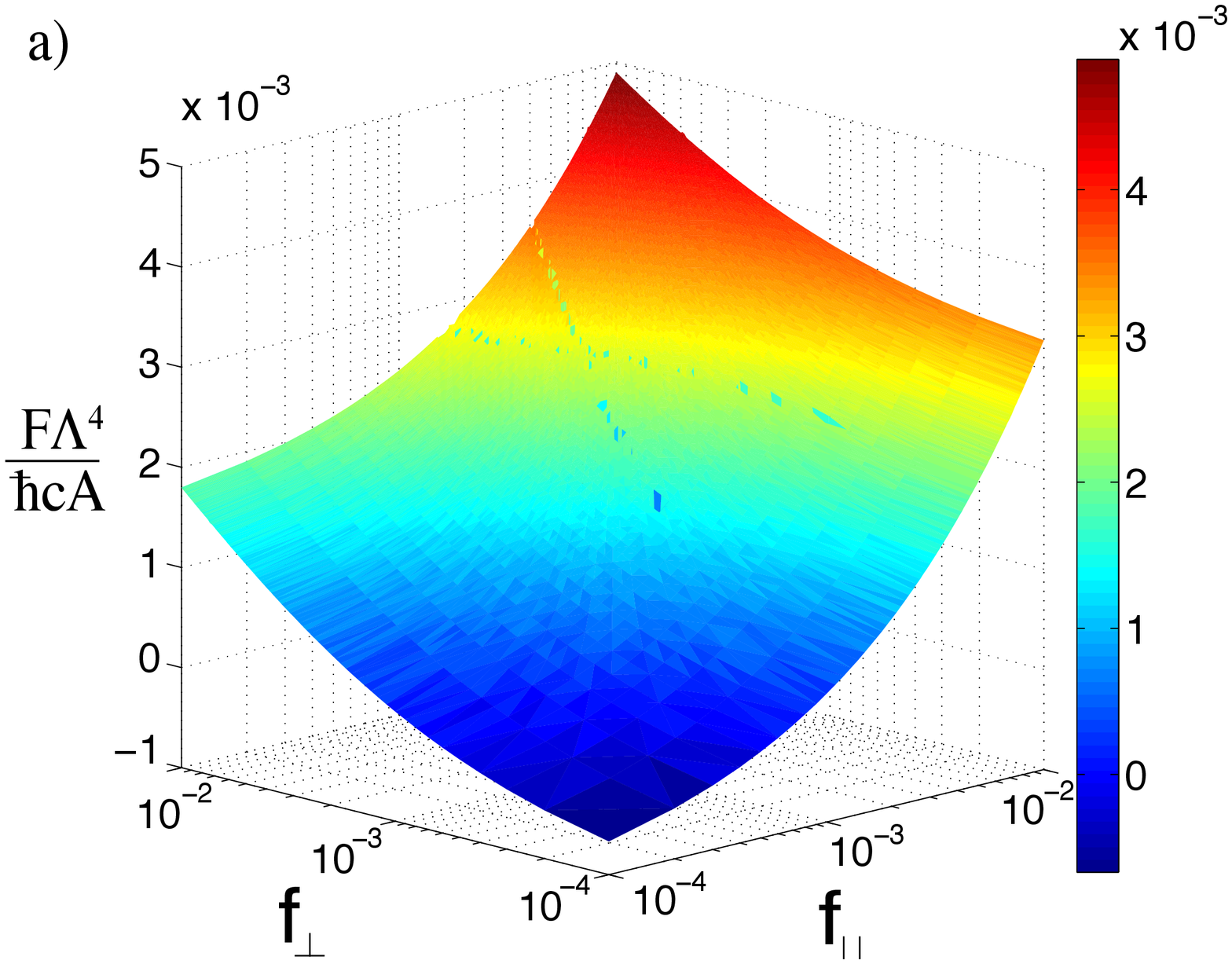}}   \\
\scalebox{0.28}{\includegraphics{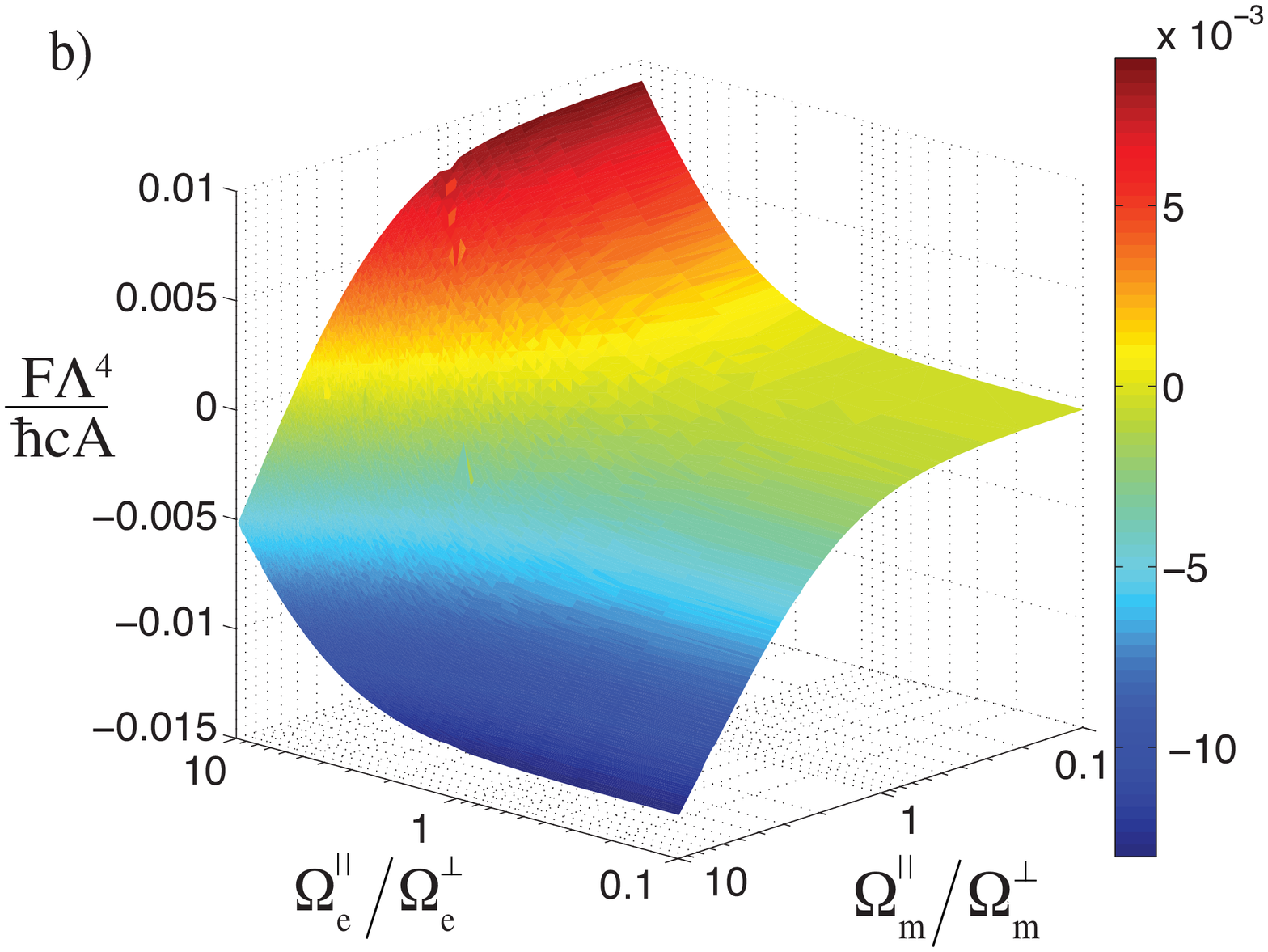}}  
\caption{Anisotropy and the Casimir force: a) effects of dielectric anisotropy for a metallic-based metamaterial
with weak Drude background; b) combined effects of dielectric and magnetic anisotropy for a
dielectric-based ($f=0$) metamaterial. The distance is fixed to $d=\Lambda$.
The perpendicular electric and magnetic oscillator strengths in b) are fixed and equal to those used in Fig. 1. All other parameters in a) and b) are the same as in Fig. 1.}
 \end{figure}

%%%%%%%%%%%%%%%%%%

{\it  Anisotropy ---} Most metamaterials have anisotropic
electric and/or magnetic activity, depending on 
the direction and polarization of the incident radiation \cite{anisotropy}. To take anisotropy into account when calculating the Casimir force we require in Eq. (\ref{eq1}) the appropriate reflection matrices. Since the general case is rather involved, we treat here a particular case in which medium 1 is isotropic and medium 2
has uniaxial anisotropy perpendicular to the interfaces
(the more general case will be treated elsewhere). 
There are several metamaterials consisting of very thin plane layers of alternating materials \cite{uniaxial}, and which could be described by permittivity and permeability tensors,
${\boldsymbol \epsilon}={\rm diag}(\epsilon_{\|}, \epsilon_{\|}, \epsilon_{\bot})$ and
${\boldsymbol \mu}={\rm diag}(\mu_{\|}, \mu_{\|}, \mu_{\bot})$, 
where $\epsilon_{\|}, \epsilon_{\bot}$ are given by (\ref{Drude Background}) 
and $\mu_{\|}, \mu_{\bot}$ by 
(\ref{Drude-Lorentz}) with respective anisotropic parameters. Anisotropy generally implies a mixing of linear polarizations, which means that the off-diagonal elements of the reflection matrix (\ref{ReflectionMatrices}) will generally not vanish.  However, the case under consideration is simple enough that we still have $ r^{sp}_2 (i \xi, {\bf k}_{\|}) =  
r^{ps}_2 (i \xi, {\bf k}_{\|}) = 0$, while the diagonal elements of the reflection matrix are \cite{Hu}
\begin{eqnarray}
\label{AnisotropicFresnelCoefficients1}
r^{ss}_2 =  \frac{\mu_{\|} K_3 - \sqrt{\frac{\mu_{\|}}{\mu_{\bot}}k_{\|}^2 + \mu_{\|} \epsilon_{\|}\xi^2/c^2}}{\mu_{\|} K_3 + \sqrt{\frac{\mu_{\|}}{\mu_{\bot}}k_{\|}^2 + \mu_{\|} \epsilon_{\|}\xi^2/c^2}}, \\
\label{AnisotropicFresnelCoefficients2}
r^{pp}_2  = \frac{\epsilon_{\|}K_3 - \sqrt{\frac{\epsilon_{\|}}{\epsilon_{\bot}}k_{\|}^2 + \mu_{\|} \epsilon_{\|} \xi^2/c^2}}{\epsilon_{\|} K_3 + \sqrt{\frac{\epsilon_{\|}}{\epsilon_{\bot}}k_{\|}^2 + \mu_{\|} \epsilon_{\|} \xi^2/c^2}} .
\end{eqnarray}

For metallic-based metamaterials with large in-plane electric response $\epsilon_{\|}(i \xi) \gg 1$
at low frequencies, it is clear from 
Eqs. (\ref{AnisotropicFresnelCoefficients1}, \ref{AnisotropicFresnelCoefficients2})
that anisotropy plays a negligible role in the determination of the reflection coefficients
when there is a dominant Drude background.
In order to better appreciate the effects 
of anisotropy we assume henceforth a small or vanishing Drude contribution. In Fig. 2a we show the Casimir force for a metamaterial that has only electric anisotropy ($\mu_{\|}=\mu_{\bot}$), which is completely coded in
different filling factors ($f_{\|} \neq f_{\bot}$). We see that a repulsive force arises only for considerably
small values of both $f_{\|}$ and $f_{\bot}$, from which we conclude that the Drude contribution must be
isotropically suppressed in order to have Casimir repulsion. Things become more interesting for 
dielectric-based ($f_{\|}= f_{\bot}=0$) metamaterials, since in that case the electric and magnetic
contributions are more similar. Accordingly, Fig. 2b shows the effect of electric and magnetic anisotropy on the Casimir force when they are present solely in terms of oscillator strengths. We chose parameters in Fig.2b  such that the force is repulsive for the isotropic case 
$\Omega_e^{\|}/\Omega_e^{\bot} =\Omega_m^{\|}/\Omega_m^{\bot} =1$. 
As expected,  repulsion is enhanced by increasing the ratio of magnetic anisotropy
$\Omega_m^{\|}/\Omega_m^{\bot}$ to the electric anisotropy $\Omega_e^{\|}/\Omega_e^{\bot}$,
while there is a crossover to attraction when such a ratio is decreased.

%%%%%%%%%%%%%%%%%%%%%%%

\begin{figure}
\begin{center}
\hspace{-20pt}
\scalebox{0.28}{\includegraphics{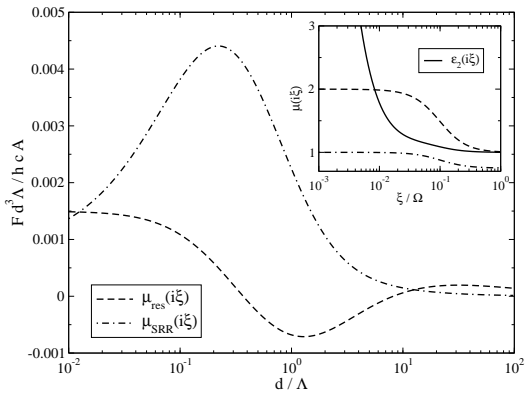}}
\vspace{-11pt}
\caption{
Casimir force between a half-space filled with gold and a metamaterial whose magnetic
response is described either by $\mu_{\rm res}(i \xi)$ or by $\mu_{\rm SRR}(i \xi)$.
The inset shows the behavior of both permeabilities as well as of the underlying permittivity
$\epsilon_2(i \xi)$. Parameters are: $f=10^{-4}$, $C=0.25$, and all the other ones are the
same as in Fig. 1.}
\label{Comparison1}
\end{center}
\end{figure}

{\it Modelling the magnetic response ---}  Having discussed how low-frequency contributions to the dielectric permittivity of a metamaterial affect the Casimir force, let us now investigate how it is affected by the magnetic permeability. Up to now we
have characterized the magnetic response of a metamaterial by
the Drude-Lorentz expression (\ref{Drude-Lorentz}), which has
been widely used in the context of Casimir physics with
metamaterials. However, it is worth pointing out that calculations based on Maxwell's equations in a
long-wavelength approximation for 
split-ring resonator (SRR) metamaterials result in a slightly different form \cite{pendry99}:
\begin{equation}
\label{Drude-Lorentz-omega2}
\mu_{\rm SRR}(\omega) = 1 - \frac{C \omega^2}{\omega^2 - \omega_{m}^2 + i \gamma_{m} \omega} ,
\end{equation}
where $C<1$ is a parameter depending on the geometry of the
split ring. The crucial difference 
between (\ref{Drude-Lorentz}) and (\ref{Drude-Lorentz-omega2})
is the $\omega^2$ factor appearing in the numerator of the latter, a consequence of Faraday's law. Although close to the resonance both expressions give almost identical behaviors, they differ in the low-frequency limit. Moreover, for imaginary 
frequencies, $\mu_{\rm res}(i \xi)>1$ while $\mu_{\rm SRR}(i \xi)<1$ for all $\xi$ \cite{passivemu}. 
Given that all passive materials
have $\epsilon(i \xi)>1$, we conclude that Casimir repulsion
is impossible for any metamaterials, such as SRRs, described approximately
by (\ref{Drude-Lorentz-omega2}), since the electric response
would always dominate the magnetic one, even without Drude background (see Fig. \ref{Comparison1}).

%%%%%%%%%%%%%%%%%%%%%%%%%%

{\it Conclusions---} 
We have considered several aspects of the Casimir force between an ordinary medium and a metamaterial.
An important conclusion of this work is that low frequencies $\xi$, less than $c/d$, make the most significant contribution to the
force (\ref{eq1}), and that the bandwidth of the dominant low $\omega$ range is much larger than the widths of
realistic material resonances. For this reason repulsive Casimir forces are possible even without
a strong magnetic response in the optical, a condition that has generally been assumed to be necessary
for Casimir repulsion at micron separations. Similarly, a sufficiently strong low-frequency magnetic response permits quantum levitation of metamaterials. This possibility applies in particular to metamaterials that are left-handed at higher frequencies, 
but, in contrast to other recent work \cite{ulf}, does not require an amplifying medium. Dissipation
and Drude backgrounds reduce the magnitude of repulsive forces or change an otherwise repulsive force to an attractive one. 
Given that Drude backgrounds are key roadblocks for 
quantum levitation, non metallic-based metamaterials
are more likely candidates to achieve the goal of
Casimir repulsion.


\begin{thebibliography}{99} 

\bibitem{reviewCasimir} For recent reviews of both theoretical and experimental work see, for instance, 
S.K. Lamoreaux, Rep. Prog. Phys. {\bf 68}, 201 (2005) and M. Bordag {\it et al.},
Phys. Rep. {\bf 353}, 1 (2001) and references therein.

\bibitem{reviewMM} Reviews of work on metamaterials have been given, for instance, by S.A. Ramakrishna, Rep. Prog. 
Phys. {\bf 68}, 449 (2005). See also the special issue IET Microwaves, Antennas and Propagation {\bf 1}, no. 1 (2007)
eds. G. Eleftheriades and Y. Vardaxoglou.

\bibitem{lif} E.M. Lifshitz, Sov. Phys. JETP {\bf 2}, 73 (1956).

\bibitem{henkel} C. Henkel and K. Joulain, Europhys. Lett. {\bf 72}, 929 (2005).

\bibitem{ulf} U. Leonhardt and T.G Philbin,  New J. Phys. {\bf 9}, 254 (2007). 

\bibitem{kats} E. I. Kats, Sov. Phys. JETP {\bf 46}, 109 (1977). See also A. Lambrecht {\it et al.}, New J. Phys. {\bf 8}, 243 (2006).

\bibitem{other} Casimir repulsion is also possible between 
non-magnetic media, as long as the inequality
$\epsilon_1(i \xi) < \epsilon_3(i \xi) < \epsilon_2(i \xi)$
holds. This is the basis for superfluid creeping out of beakers,
and for frictionless bearings. See
F. Capasso {\it et al.}, IEEE J. Select Topics Quantum Electron. {\bf 13}, 400 (2007) and references therein.

\bibitem{boyer} T.H. Boyer, Phys. Rev. A {\bf 9}, 2078 (1974).

\bibitem{klich} O. Kenneth {\it et al.}, Phys. Rev. Lett. {\bf 89}, 033001 (2002).

\bibitem{capassocomment} D. Iannuzzi and F. Capasso, Phys. Rev. Lett. {\bf 91}, 029101 (2003).

\bibitem{opticalMM} V.M. Shalaev, Nature Photonics {\bf 1}, 41 (2007). 

\bibitem{activeMM} In fact in the recent paper by C. Raabe and D.-K. Welsch, arXiv:0710.2867v1, it is argued for just these same reasons that the prediction of a repulsive force in Ref. \cite{ulf} is incorrect.

\bibitem{CapassoHSM} D. Iannuzzi {\it et al.}, Proc. Nat. Acad. Sci. USA {\bf 101}, 4019 (2004).

\bibitem{irina} I.G. Pirozhenko and A. Lambrecht, arXiv:0801.3223.

\bibitem{photoniccrystals}
K.C. Huang {\it et al.}, 
Appl. Phys. Lett. {\bf 85}, 543 (2004).

\bibitem{dielectrics} J.A. Schuller {\it et al.}, Phys. Rev. Lett. {\bf 99}, 107401 (2007).

\bibitem{anisotropy} See, for instance, 
S. Linden {\it et al.}, Science {\bf 306}, 1351 (2004);
A.V. Kildishev {\it et al.}, J. Opt. Soc. Am. {\bf B 23}, 423 (2006).

\bibitem{uniaxial} See, for instance, J. Schilling, Phys. Rev. {\bf E 74}, 046618 (2006);
T. Tanaka {\it et al.}, Phys. Rev. B {\bf 73},  125423 (2006).

\bibitem{Hu} See, for instance, L. Hu and S.T. Chui, Phys. Rev. B {\bf 66} 085108 (2002).

\bibitem{pendry99} J. B. Pendry {\it et al.}, IEEE Trans. Microwave Theory Tech. {\bf 47}, 2075 (1999).
High-frequency corrections to this formula that would ensure $\mu_{\rm SRR}(\infty)=1$ are not found to
affect our numerical results or conclusions.             

\bibitem{passivemu} Unlike $\epsilon(i \xi)$, $\mu(i \xi)$
can be less than $1$ without violating Kramers-Kronig relations. See, for instance, L.D. Landau, E.M. Lifshitz and L.P. Pitaevskii, {\it Electrodynamics of Continuous Media} (Elsevier, Oxford, 2007).



\end{thebibliography}
\end{document}